\documentclass[journal]{IEEEtran}
\IEEEoverridecommandlockouts

\usepackage{cite}
\usepackage{amsmath,amssymb,amsfonts}
\usepackage{graphicx}
\usepackage{booktabs}
\usepackage{subcaption, color}
\usepackage[inline]{enumitem}
\usepackage{balance}

\begin{document}
\title{QoE Evaluation for Adaptive Video Streaming: Enhanced MDT with Deep Learning}

\author{Hakan Gokcesu, Ozgur Ercetin, Gokhan Kalem, Salih Erg\"{u}t%
\thanks{H.~Gokcesu is with Department of Electrical and Electronics Engineering, Bilkent University, Ankara, Turkey.}
\thanks{O.~Ercetin is with the Faculty of Engineering and Natural Sciences, Sabanci University, Istanbul, Turkey. }
\thanks{G.~Kalem and S.~Erg\"{u}t are with the Department of 5G Research and Development, Turkcell Technology, Istanbul, Turkey. }
\thanks{This work is supported in part by The Scientific and Technological Research Council of Turkey (TÜBİTAK), funded by under 1001 Programme Project No: 119E198, and ``5G-PERFECTA: 5G and Next Generation Mobile Performance Compliance Testing Assurance” under EUREKA Programme, funded by under TEYDEB 1509 - International Industry R\&D Support Program, Project No: 9190006. H.~Gokcesu is supported by Turkcell within the framework of 5G and Beyond Joint Graduate Support Programme coordinated by Information and Communication Technologies Authority in Turkey.}
}

\maketitle

\begin{abstract}
The network performance is usually assessed by drive tests, where teams of people with specially equipped vehicles physically drive out to test various locations throughout a radio network. However, intelligent and autonomous troubleshooting is considered a crucial enabler for 5G- and 6G-networks. 
In this paper, we propose an architecture for performing virtual drive tests by facilitating radio-quality data from the user equipment. Our architecture comprises three main components: i) a pattern recognizer that learns a typical pattern for the application from application Key Performance Indicators (KPI); ii) a predictor for mapping network KPI with the application KPI; iii) an anomaly detector that compares the predicted application performance with that of the typical application pattern. In this work, we use a commercial state-of-the-art network optimization tool to collect network and application KPI at different geographical locations and at various times of the day for training an initial learning model. 
We perform extensive numerical analysis to demonstrate key parameters impacting correct video quality prediction and anomaly detection. We show that the playback time is the single most important parameter affecting the video quality, since video packets are usually buffered ahead of time during the playback. However, radio frequency (RF) performance indicators characterizing the quality of the cellular connection improve the QoE estimation in exceptional cases. We demonstrate the efficacy of our approach by showing that the mean maximum F1-score of our method is 77\%. Finally, the proposed architecture is flexible and autonomous, and it can operate with different user applications as long as the relevant user-based traces are available. 
\end{abstract}


\section{Introduction}

\noindent With the evolution of cellular networks, there has also come an increase in structural complexity and heterogeneity of services, requiring constant monitoring of the communication systems. This landscape will only become more complex with the evolution of 5G and the promise of 6G on the horizon. Most importantly, new systems will require intelligent and automated troubleshooting tools, which is the focus of our work.
 
Network operators put significant effort into network planning, monitoring, and optimization to deliver a satisfactory customer experience with provided services. In addition to the continuous data collection from network monitoring probes, they regularly perform drive tests for capacity and coverage optimization and quality of service monitoring. 

Drive-tests are usually performed with commercial-off-the-shelf testing equipment (such as Keysight NEMO, Infovista TEMS, Rohde-Schwarz SwissQual) capable of running a predefined set of tests and collecting relevant Key Performance Indicators (KPIs) on the radio, core, and transport network protocols. In addition, device-level and application-level statistics are also recorded during these drive tests. The aim of drive-tests is to model the user experience more realistically; however, this has become even more challenging as the networks became more complex with added multi-layered heterogeneity as they have evolved from 2G to 5G. 

Since such drive tests are costly due to the need for human resources and vehicles traveling long distances, a considerable amount of time passes between successive drive tests and only covers a small part of the customer base. Therefore, they can only provide delayed and crude perceptions of the network performance. For this reason, 3GPP has been publishing technical specifications on the minimization of drive-tests (MDT) since Release 10 (3G) in 2010, with the latest version\footnote{3GPP TS 37.320: 
``Radio measurement collection for Minimization of Drive Tests (MDT); Overall description; Stage 2 (Release 16)"} covering 5G networks in Release 16, to facilitate remote collection of radio quality, quality of service and geolocation data from the user equipment. 

Today, video streaming (live and on-demand) platforms play a significant role in consuming the highest data volume in cellular networks. Based on the operator statistics under this research, YouTube mobile data volume corresponds to approximately 60\% of the total data volume. As understood, in addition to its on-demand and live-streaming enriched video contents, the YouTube application constitutes a significant part of mobile data consumption across the network with the help of various resolution levels (up to 4K). Moreover, according to ``similarweb.com'', it is the highest ranking TV Movies And Streaming platform worldwide. All of these build a real motivation to consider YouTube in this work. 

There are several ways to estimate the quality of such services. Traditionally, the methods to gain insight into the delivered quality of service and the users' experience have been through controlled laboratory experiments, where users' opinions have been collected. The results are then reported in Mean Opinion Scores (MOS), corresponding to the average of users' opinions. These techniques are often referred to as subjective quality assessment. There are reliable methods standardized by The International Telecommunication Union - Telecommunication Standardization Sector (ITU-T) study groups to conduct them in various industry platforms. The objective way of testing uses an image-based quality metric based on fully decoded video output or bitstream-based metrics that evaluate the IP traffic and headers only as defined in the video quality estimation module; ITU-T P.1203.1 Recommendation. Originally designed as the output of subjective tests where the subjects are asked to rate their satisfaction over a scale of 1 (`Bad') to 5 (`Excellent'), MOS is also the output of the objective model given in ITU-T P.1203 Recommendation. This up-to-date version of the Recommendation is built on previous versions of MOS calculation models in the literature \footnote{Recommendation ITU-T P.1203.1 (2016), Parametric bitstream-based quality assessment of progressive download and adaptive audiovisual streaming services over reliable transport – Video quality estimation module.}. 

Network services have gained significant intelligence with the advent of 5G, involving intensive access to data communication and computing resources. However, although quality assurance in cellular networks has been studied extensively in the last decade, there is still a lack of understanding of how network KPIs affect the application KPIs and, eventually, Quality of Experience (QoE). In particular, quality assurance for video streaming services is complex due to the technological heterogeneity in both network infrastructure end-to-end and user equipment side, including applications running on Android, iOS, or the HarmonyOS platforms. Additionally, the operators rarely have access to the application level KPIs via network probes or monitoring tools, so resorting to costly drive-tests, as mentioned earlier. 

Addressing these challenges requires novel data-driven approaches to interpret monitored data and observations correlated from the network and application performances. 
To this end, our contributions in this work are as follows:
\begin{itemize}
    \item We analyze drive test user traces from a leading cellular network operator in Turkey collected from different geographical locations and at various times.
    \item We use time-series user data to train a Recurrent Neural Network autoencoder to learn the typical MOS pattern for the reference video.
    \item We use tree-ensembles-based machine learning models to predict the MOS of a benchmark video based only on the radio KPIs monitored by the base station. We show that our prediction method has a mean square of error (MSE) of approximately 3\% of the highest MOS.
    \item We use decision trees and What-if analysis tools, i.e., SHapley Additive exPlanations (SHAP), for interpreting the results.
    \item We identify the video playback time as the most crucial feature for MOS prediction, which confirms the buffering effect in adaptive streaming.
    \item We present an unsupervised anomaly detection technique that classifies a session anomaly if the predicted MOS deviates from the typical MOS pattern by a significant amount.
    \item We demonstrate that although MOS has the highest dependency on playback time, the prediction of MOS improves by considering network KPIs during anomalies.
\end{itemize}
Note that although our architecture is sufficiently general to be applicable to any user application on a smartphone, we demonstrate its performance with YouTube as the leading adaptive video streaming application since the QoE metric is well defined, and it also constitutes a huge data volume for cellular networks.

The paper is organized as follows. We first review the relevant literature in Section \ref{sec:related_work}. In Section \ref{sec:qoe}, we define QoE and its measurement for adaptive video streaming. Section \ref{sec:model} describes our architecture and data collected from drive tests. In Section \ref{sec:MOS_pattern} and \ref{sec:MOS_Pred}, we explain the pattern recognition and QoE estimation components of the architecture. We explain the anomaly detection mechanism in Section \ref{sec:anomaly}. Section \ref{sec:interpret}, we discuss the results of our experiments to identify the importance of features and interpret the decisions of the QoE benchmark predictor. We verify Signal-to-Noise-Ratio (SNR) as the key KPI indicator for an anomaly in Section \ref{sec:root}. Finally, we provide our conclusions and future directions in Section \ref{sec:conc}.

\section{Related Work}
\label{sec:related_work}
\noindent Unlike the measurements for Quality of Service (QoS), which are the network-related performance indicators, QoE quantities measure the satisfaction of end-users, who are the decision makers in the era of 5G \cite{Tsolkas}. The prediction of QoE have been a subject of research since 3G networks era \cite{casas2013youqmon}.
Estimation of the variables affecting QoE has been recognized as an important problem for cloud service providers \cite{Mushtaq}, and video streaming services \cite{Bouraqia}, e.g. IPTV \cite{frnda2019hybrid,JongKim} and YouTube \cite{orsolic2016youtube, orsolic2018youtube}. With the advent of 5G, QoE modeling has been heavily investigated in high resolution/bit-rate services such as Ultra HD video streaming \cite{Nightingale2018} and remote control via virtual reality simulators \cite{Brunnstrom2020}. Some aim of the works investigating QoE included determining how certain network-wide schemes should be realized, such as adaptive modifications of bit-rate \cite{Ibarrola} and media playback speed \cite{Perez2019}. Similarly, there were also works that construct databases which correlate QoE with the video content, network conditions, and adaptive bit-rate schemes \cite{bampis2018perceptually}. Hence, there are a plethora of works that investigate and scrutinize QoE estimation.

There are many aspects affecting QoE quantities \cite{Bouraqia}. Popular approaches in QoE estimation incorporate QoS measurements \cite{Msakni}, packet information or even the media itself \cite{Yamagishi, frnda2019hybrid}. In one example, \cite{34} investigates the effects of stall, initial delay, and visual quality for end-user QoE. There is also a protocol-based approach, for HTTP adaptive streaming \cite{Robitza} as well as an approach called `affective computing', which tracks the end-users through their facial expressions \cite{Amour}. Similarly, \cite{6} has also proposed alternative means to cope with complex QoS and QoE mapping.

In our models, the specific QoS metrics in our consideration are related only to RF signal quality, unlike those models involving packet loss, delays, throughput, and the like, as in \cite{Miguel,pokhrel2013estimation,casas2013youqmon,kang2013artificial,Brunnstrom2020,Mushtaq,Pierucci16,Banovic-Curguz,Pierucci15}, and application-layer KPIs as in \cite{orsolic2016youtube,kang2013artificial,Uthansakul}, all of which would require access to user traces. Although there have been some conclusions that QoS measurements alone are insufficient to estimate QoE \cite{Msakni}, it is possible to perform predictions necessary for handling network related causes of QoE degradation, as shown in our work. Ideally, it may be desirable to obtain formula-based parametric QoE estimators; however, in recent works, such a venture proved to be challenging and convoluted, even for the same service type \cite{Tsolkas}.

In the endeavor of QoE prediction, systematically quantifying QoE variables, was also investigated in detail. \cite{18} reviews some recent methodologies to predict QoE. The proper metrics for QoE, along with their modeling, are by themselves a topic of interest in the literature \cite{liotou2016roadmap}. Note that, in our work, we do not investigate the problem of subjective QoE labeling, which by itself is also a topic of interest \cite{Chang}. Our collection of QoE measurements are mainly generated by PEVQ-S algorithm exploited in the drive-tests. 

In our work, we attempt to predict the MOS, using network KPIs, namely Reference Signal Receive Power and Quality (RSRP and RSRQ), and Received Signal-to-Noise Ratio (SNR). The motivation behind using network KPIs is the fact that QoE is primarily affected by network conditions which may lead to packet losses, data-rate reduction, and delays \cite{Yamagishi}. Following that, we classify whether the current state of QoE is anomalous in an unsupervised manner. We employ machine learning (ML) and artificial intelligence (AI) approaches, similar to \cite{pokhrel2013estimation,Ibarrola} and \cite{kang2013artificial,Pierucci16,Uthansakul,Bampis,Pierucci15}. Unlike \cite{Ibarrola}, we are restricting ourselves to only working with device-based RF performance indicators. Dimopoulos et al.,\cite{11} also leverage ML to evaluate the correlation between QoS and QoE to overcome the challenge of measuring end-user satisfaction. Another area where ML/AI techniques were recently shown to have a lot of promise is improving and designing Adaptive BitRate (ABR) schemes. The authors in \cite{2,10} use ML to compute parameters of the existing ABR scheme to adapt to dynamically changing network conditions. In \cite{Liu}, the authors use deep learning AI for content-aware ABR via QoE estimations. Our work is also similarly adaptable for real-time/online learning and testing in dynamic environments via iterative processing and, unlike aforementioned and \cite{menkovski2010online, orsolic2018youtube,Batalla2018,Bampis}, we do not require the incorporation of throughput or content-related quantities. 
 
In comparison to the literature, our work considers a regression ML model that only incorporates last-mile RF performance indicators such as SNR, RSRP/Q, and PRB as input to map QoS to QoE. Also, note that our ML model is unaware of the underlying algorithm type (and its characteristics) used for streaming the adaptive video and mandates reasonable realistic results.



\section{Quality of Experience of Adaptive Video Streaming}
\label{sec:qoe}
\noindent Quality of Experience (QoE) is defined as the overall acceptability of an application or service, as the end-user perceives. QoE has become the fundamental metric in evaluating user satisfaction with the advent of LTE Advanced and 5G. However, QoE presents several new challenges for the operators. 
For one, subjective tests require the participation of many end-users in controlled experiments, so it is not practical to conduct them frequently. Hence, for day-to-day operations, operators resort to objective QoE measurements to approximate the results of the subjective tests. 
 
 

HAS-type players pull video segments from the server independently and maintain the playback session states locally so that the servers do not need to track individual session states. The structure of each video stream is described in a Media Presentation Description (MPD) file. When a video client wants to play a specific video stream, the client first downloads the MPD file. Then, the client’s video adaptation algorithm determines the video bit-rate of the next segment and the target inter-request time. The player also tracks the time it takes to download each segment, and if the download duration is shorter than the target inter-request time, the client experiences an off time. The downloaded segments are queued on the player’s buffer and are de-queued during streaming. Incorrect choice in representation bit-rate can cause the end-user to experience a series of stall events. Stalling describes video playback interruption due to buffer under-run. When the buffer level is beneath a given threshold, insufficient data for playback is available, and the video playback has to be stopped until the buffer is refilled.
 
 
QoE depends on the network conditions and QoS metrics and the players' buffering approaches and adaptation capabilities. These application-based approaches are also affected by the algorithm changes in the mobile operating systems, e.g., Android and iOS. Furthermore, network operators employ video-centric data optimization tools between the core and packet data networks to make mobile video services resilient and efficient in terms of network resources and quality management. The effect of radio network KPIs on the MOS is less pronounced due to these enhancements affecting buffering and total experience at the user side.

\section{Architecture}
\label{sec:model}
\begin{figure*}
    \centering
    \includegraphics[width=0.7\linewidth]{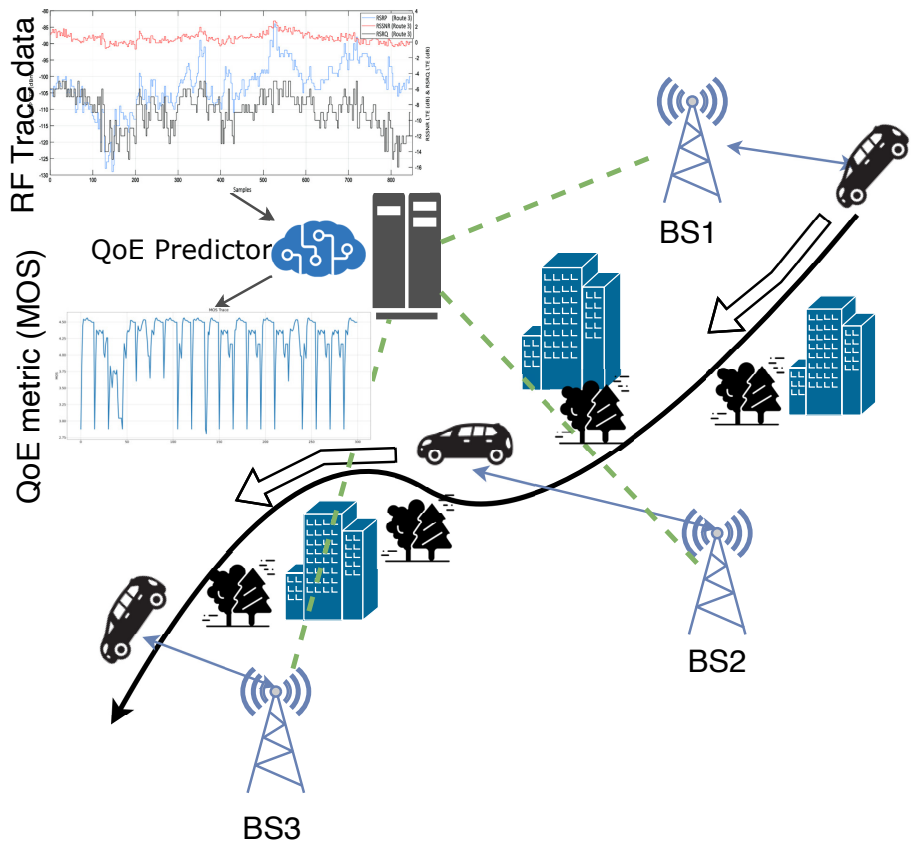}
    \caption{Proposed Base Station in-the-loop Virtual Drive Test Architecture. A mobile user moves from the right to the left of the figure, and its connection is handled by three base stations. The back-end server collects trace data from the base stations and feeds it to the deep learning module to evaluate the QoE for this user acting as a virtual drive-test vehicle. }
    \label{fig:model}
\end{figure*}
\subsection{Overview}
\noindent User equipment-based Quality of Service parameters are measured in a specific geographical area such as a base station site, cluster, or a driving route with a network trace mechanism following 3GPP technical specifications. The trace mechanism is run on operations support systems (OSS) which comprise network elements of 3G, 4G, and 5G. Nevertheless, this operation is executed for a limited duration (e.g., a couple of hours when desired) since it creates extra processing in the nodes. Then, subscriber-specific data is masked and not disclosed within the dataset because of its privacy. Meanwhile, this trace mechanism additionally lacks the capability to interpret QoE to reflect the customer satisfaction on an application. 

Predicting the MOS of an adaptive video streaming session involves many challenges. The most critical challenge is monitoring and collecting application KPIs, which is the primary input of objective MOS evaluation. Therefore, the network operator requires an effective technique to map network KPIs to application KPIs for accurate MOS prediction. In this work, we use machine learning techniques to accurately predict user experience in a video platform from network KPIs. Another vital challenge is identifying the \textit{anomalies} in the video platform. Note that an inferior instantaneous network KPI does not always correspond to a poor application KPI, thanks to the buffering of video data. Hence, prolonged exposure to bad network conditions causes poor user experience. 

Our methodology is to find and interpret anomalies in the last mile of a cellular network in an unsupervised way. In unsupervised learning, there are no labeled examples to use. To understand whether the user experience is poor or not, we first need to know the typical time-series pattern of the application behavior. We use a recurrent neural network (RNN) auto-encoder to recognize the reference video's typical MOS pattern. A session with a predicted MOS pattern substantially deviating from the typical MOS pattern is identified as an anomaly. 

This typical time-series pattern of MOS is intended to represent the de-noised MOS values over time under the current viewing experience and its factors. Since the experiments of this work are done in succession, this pattern is presumably generated for the non-changing factors such as the overall base KPI performances (not the minute changes but the general performance) and the specifics of the streaming video (which is the same for each viewing experience of the same content). In practice, although generating patterns for different base KPI performances seems suitable, the generation of such patterns for every video specifically may not be practical. However, note that, since this pattern extraction analyzes the MOS values in a viewing by aggregating several realized viewings, one can potentially achieve it by using samples across different videos via first clustering the realized viewings and then aggregating each video cluster separately. In the end, a resulting pattern would not only be representative of a single video but a number of similar videos. Nevertheless, the commercial-off-the-shelf measurement tools all use a reference YouTube video as a benchmark.

Figure \ref{fig:model} summarizes our proposed architecture. We first learn the typical behavior of the application to determine what to expect on the end-user side. Then, we predict the MOS of the benchmark video by only exploiting some major network KPIs selected based on our correlation analyses. Finally, we compare our predicted MOS values with the typical MOS pattern to find out anomalies in an unsupervised manner. 

\begin{figure}
	\centering
	\includegraphics[width=0.8\linewidth]{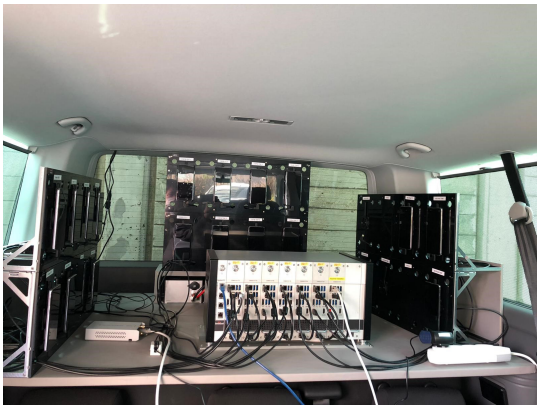}
	\caption{Measurement Platform in the Drive-Test Vehicle.}
	\label{fig:test vehicle}
\end{figure}
\begin{table}[]
    \centering
        \caption{YouTube Application KPIs collected by the measurement platform.}
    \begin{tabular}{|c|c|}
    \hline
          \textbf{YouTube Key Performance Indicators}& \textbf{Units} \\
          \hline
         Video Request Success Rate& (\%)\\
Video Access Time & seconds\\
Video Initial Buffering Time & seconds\\
Video Stalling Rate & (\%)\\
Video MOS& 1-5 \\
Average Resolution & pixels\\
Video Completion Rate & (\%)\\
Successful Session& Number of\\
Session Drop & Number of\\
Average Application Throughput & Mbps\\
\hline
    \end{tabular}
    \label{tab:KPI_list}
\end{table}
\begin{figure}
	\centering
	\includegraphics[width=1.0\linewidth]{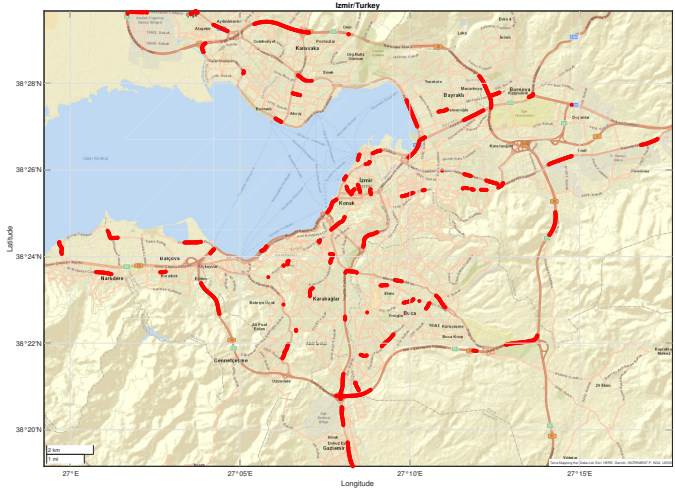}
	\caption{Red points on the map show the locations of measurements made during a drive-test in Izmir-the third largest city in Turkey.}
	\label{fig:video mos}
\end{figure}
\subsection{Data Collection and Analysis}
\noindent The data is collected by a drive-test vehicle over pre-determined routes in the predominantly urban areas, including dense-urban and sub-urban districts. The routes are selected to consider different radio network infrastructures (i.e., different network vendors) and conditions. Moreover, the test routes include main boulevards and highways and side streets with extensive attenuation and shadowing due to concrete buildings in a dense type of urbanization. The attention to the drive-test routes helps us to better model the variable QoE and service characteristics resulting from the dynamic and heterogeneous structure of the network. The drive-tests are predominantly run during the daytime, mostly between 08.00 - 19.00, in road traffic conditions common for the time of the day in that city. Fig. \ref{fig:test vehicle} demonstrates the measurement platform, mounted at the backside of the drive-test vehicle (i.e., minivan), consisting of smartphones, testing equipment, antennas and connection cables placed on the desk, which also has benchmarking capabilities (e.g., comparing networks, apps, devices, data and voice services, etc.). 

We demonstrate our methodology by an analysis of the YouTube application. At the time of this writing, YouTube stands out as the second mobile application in the ranking according to total data traffic in the operator network, right after Instagram. Requirement analysis is carried out, and YouTube application-based performance metrics that are critical for understanding adaptive video streaming experience are identified by considering measurement tool capabilities. These application KPIs are indicated with their measurement units in Table \ref{tab:KPI_list}. 
 
During the drive-test, a reference YouTube video\footnote{https://youtu.be/A1OpjNq6zGE}, recommended by the measurement tool vendors (e.g., Keysight, Accuver, etc.), is selected and executed in the test script. Based on the test scenario determined by the operator, the duration of video downloaded is limited to 60-second sessions; however, the downloaded content in this period is not fixed due to the varying video resolutions in every session. \textbf{The tests are sharply stopped at the 60 second-mark even though the reference video is approximately 4.5 minutes long.} This duration is regarded as sufficient to assess the video quality due to the playback buffer mitigating the impacts of RF deterioration in the long run. On the contrary, when the playback buffer runs empty due to congested or poor network conditions, adaptive streaming encounters a stalling event as soon as the network cannot handle the timely arrival of sequential packets. Therefore, every 60 seconds of the test measurement is identified as a single session. Fig. \ref{fig:video mos} displays the drive-test routes in a dense urban area in the city of Izmir. Most measurements on this route have MOS$>4.1$, demonstrating that the network conditions in urban areas are highly optimized. However, this also causes problems for autonomous troubleshooting since the data for inferior conditions is rare, and these cases cannot be learned effectively by machine learning techniques.

\begin{table}[]
    \centering
        \caption{Network KPIs and the number of samples in test data.}
    \begin{tabular}{c|c}
         \textbf{Network KPIs} & \textbf{Number of Samples} \\
         \hline
         RSRP & 84296\\
         RSRQ & 84296 \\
         SNR & 116080 \\
         PRB (PDSCH) & 49680
    \end{tabular}
    \label{tab:net_KPI}
\end{table}
\begin{table}[]
    \centering
    \caption{Application KPIs and the number of samples in test data.}
    \begin{tabular}{c|c}
        \textbf{Application KPIs} &\textbf{ Number of Samples} \\
        \hline
         Video MOS & 16328\\
         Video Access Time & 78464\\
         Video Initial Buffering Time & 78296\\
         App Throughput (DL) & 34312 \\
         Streaming Resolution & 1128112 \\
         Video Streaming State & 77400
    \end{tabular}
    \label{tab:app_KPI}
\end{table}
\subsection{Data Processing}

\noindent The measurement tool used in drive-tests, fully compatible with Android-based smartphones, comprises software and scalable hardware that can handle different test scenarios. It supports comprehensive application tests such as YouTube, Instagram, Facebook, Twitter, and the ability to collect and visualize real-time network KPIs measured over connected devices throughout the route. In addition, this testing platform provides network QoS measurements and a variety of application testing, including voice and video streaming quality measurement algorithms such as PESQ, POLQA, and PEVQ-S based on ITU-T standards. The dataset used in this paper is generated by Opticom's implementation of PEVQ-S. Note that PEVQ-S is the most common algorithm used by operators worldwide \cite{keyhl2014perceptual}.
In this work, the PEVQ-S algorithm, developed to analyze the perceptual evaluation of streaming video quality and support full scalability from high to low video bit-rates and screen sizes, is employed to measure video MOS for the reference YouTube video. 

The measurement tool periodically generates video MOS estimates every 4-5 seconds and outputs Radio Frequency (RF) measurements every second. Nevertheless, some sessions may have none, a few seconds, or more than two minutes of measurement data due to software or test vehicle glitches. Hence, data cleansing is performed to keep only YouTube application KPIs and remove corrupt measurements. 
The resulting dataset has a single row per RF value (e.g., RSRP, SINR, etc.) and event (e.g., session start, stall, etc.), as well as hundreds of columns summarizing all drive-test dimensions (such as date, time, longitude, latitude, session number, cell ID info, channel number) and network and application KPIs related to that particular test session of YouTube video. In this context, we have processed a comprehensive dataset comprising roughly 1.6 million rows and 152 columns in total, collected from drive-tests performed in the three most populated cities in Turkey; Istanbul, Ankara, Izmir. Precisely, our dataset consists of 1548 video test sessions\footnote{Each test session is at most 60 seconds long of measurement of radio and application KPIs collected for the same Youtube video.}, and 1199 of these have at least 12 MOS measurements in each. 

The MOS values are measured during a test session (i.e., a single viewing) periodically in a non-strict manner such that each viewing may have a different number of MOS values measured depending on external factors. Note that, these MOS values are different from the traditional (and literal) sense of asking the opinions of real people regarding their viewing experience. The MOS values we use are standardized mechanical estimates which take into account many aspects such as the quality of a video, stalls, among others.

In summary, Tables \ref{tab:net_KPI} and \ref{tab:app_KPI} indicate the number of samples for each network and application KPI, respectively utilized in the post-processing phase of this work.

Major network KPIs are selected based on correlation analysis, and they are explained as follows: 
\begin{itemize}
\item \textit{Reference Signal Received Power (RSRP)} is the average power received from a single reference signal, and its typical range is around -44 dBm to -140 dBm\footnote{dBm indicates power level expressed in decibels with reference to one milliwatt}. In practice, RSRP levels for usable signal typically range from about -75 dBm close to an LTE cell site to -120 dBm at the edge of LTE coverage.
\item \textit{Reference Signal Received Quality (RSRQ)} indicates the quality of the received signal, and its reporting range is between -3 dB (decibels) and -19.5 dB. RSRQ measurement on user equipment provides additional information when RSRP is insufficient to make a reliable handover or cell re-selection decision in the network. 
\item \textit{Signal-to-Noise Ratio (SNR)} is a measure of signal quality as it quantifies the relationship between RF conditions and user throughput. Thus, it is a common practice to use SNR as an indicator for network quality. 
\item \textit{Physical Resource Block (PRB)} consists of 12 consecutive sub-carriers for one slot (0.5 ms). It is the smallest element of resource allocation assigned by the base station scheduler in the radio network. Radio resources may be defined in different ways, but it is convenient to choose PRB as the resource being shared in the case of LTE network. 
\end{itemize}

\textbf{Since these RF KPIs can also be collected on the network side using call traces, any model created based on them can be extended to be used for network-based QoE estimation.} As specified by 3GPP 32.421, subscriber and equipment trace provide very detailed information on RF coverage through performance measurements at call level when enabled for specific mobile equipment to allow remote monitoring and optimization operations \cite{3gpp.32.421}. As opposed to permanent performance measurements, such traces are selectively activated for a limited period of time.

\textbf{Note that the throughput, which is a principle KPI indicating the network's quality of service, is not considered herein. }The underlying reason is that application throughput, a meaningful KPI to utilize in such application-based experience, can only be collected by drive-test equipment in the field. In contrast, the other mentioned KPIs are easily reachable in the statistics database of the network and do not require extra effort, time, and cost. Moreover, the measurement of throughput in the cellular network is primarily based on the user device. It corresponds to the aggregated throughput for data transmitted over the user device and not for a specific application running on that device. Hence, aggregated throughput measured at the network side may include data downloaded by other applications running on the same device, and it cannot reflect correct application-based experience.

\begin{figure*}
    \centering
    \includegraphics[width=1.0\textwidth]{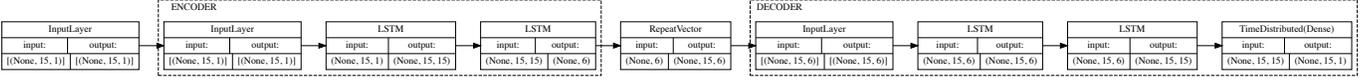}
    \caption{Autoencoder model for learning the typical MOS pattern.}
    \label{fig:NN_typical_MOS}
\end{figure*}


\begin{figure}
    \centering
    \includegraphics[width=0.4\textwidth]{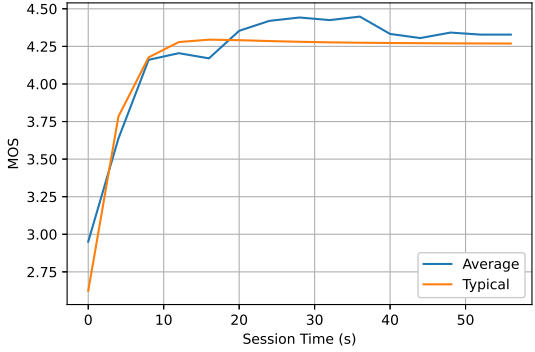}
    \caption{Typical and average MOS patterns learned from autoencoder given in Fig \ref{fig:NN_typical_MOS} with training ratio taken as 75\%.}
    \label{fig:MOS_pattern}
\end{figure}
\section{MOS Pattern Recognition}
\label{sec:MOS_pattern}
\noindent One of the major challenges in network maintenance is to track the application-specific dynamic behavior for streaming video, which may not reflect anomalies in the network but instead the characteristics of the application. Video streaming is designed to face different network conditions during a video session, making the rate adaptation algorithm switch requested segment qualities to adapt fast to the changes. For example, HAS applications differ according to initial buffering period, video access time, rate adaptation algorithm, behavior when stalling occurs, and others. These factors jointly affect the user experience, which is quantified as MOS. 

In order to {\em learn} the streaming application behavior, we use time-series MOS data available for each video session. The family of Recurrent Neural Networks (RNNs) can learn based on the time sequence by specifying hidden states that depend not only on the input but also on the previous hidden state. The most important characteristic of RNNs is using memory to process sequences of inputs. It means, unlike the convolutional neural networks (CNN) architecture where there is no memory to capture the temporal correlation, RNNs have the capabilities to capture the sequential correlations and information. In this work, we use Long Short Term Memory (LSTM) RNN to learn the typical MOS pattern for the test application and video. In the recent decade, LSTM has become the most popular RNN used for time-series prediction because of its controlled exposure of the memory content and its computation of the new memory content without any separate control of the amount of information flowing from the previous time step.

Note that this pattern recognition is not intended for use in predicting MOS values. It is the extraction of general session MOS behavior from a multitude of sample sessions. In a way, it is a robust aggregation approach.

Fig \ref{fig:NN_typical_MOS} represents the LSTM RNN \textit{autoencoder} architecture adopted for MOS pattern recognition. Autoencoder is an unsupervised learning technique where there is a bottleneck in the network which forces a compressed knowledge representation of the original input. If there are dependencies between the input features 
, this structure can be learned and leveraged when forcing the input through the neural network's bottleneck. The autoencoder forces the model to maintain only the variations in the data required to reconstruct the input without holding on to redundancies within the input.
We employ autoencoder-based learning to remove the noise in MOS measurements inherent due to varying network conditions and measurement tool inaccuracies. 

The collected MOS values (measurements) are taken as the ground truth. However, they do not well-represent the qualities of a viewing resulting from a video itself under fast-changing dynamics. To circumvent that, an autoencoder is utilized to stabilize the MOS values, i.e. the discrepancies resulting from changing performance indicators are somewhat treated as noise. What remains is a typical MOS value representation for a given video and the overall network conditions. The effects of sudden or temporary changes in the network conditions on the MOS measurements are thus eliminated.

Note that the typical MOS pattern obtained should mainly reflect the \textit{ideal} rate adaptation characteristic of the streaming application only. Fig. \ref{fig:MOS_pattern} illustrates what we mean as the typical MOS pattern learned and how it compares with the MOS pattern obtained by simply averaging over all sessions. The average MOS reflects fluctuations which are most probably due to the high variability in network conditions. On the contrary, typical MOS reflects the application behavior, which begins by requesting low video resolution and quickly ramping up when the network throughput is sufficient.

In the following, we use the term \textbf{\em{feature}} to refer to features in machine learning, which are not to be confused with features in any other context. Features in machine learning are akin to explanatory variables used in statistical analysis. They come in the form of measurable properties or descriptive characteristics, such as network KPI measurements. They refer to the input segments of the data samples at hand instead of the dependent variables or the output segments, i.e., the targets of predictions (e.g., MOS values).

The autoencoder, as depicted in Fig \ref{fig:NN_typical_MOS}, takes 
at most 15 MOS values per session as input features to train an LSTM layer with 15 outputs which are then inputted to a hidden LSTM layer with six outputs, where each output approximately represents 10 second time blocks. Hence, we aim to remove the noisy measurements and find the essential pattern of the MOS. The decoder is the opposite of the encoder, with time distributed dense layer giving back the 15 features. We choose Rectified Linear Unit (ReLu) as the activation function, Mean Square Error (MSE) as the loss function, and Adam\footnote{Adam is a stochastic gradient method that is based on adaptive estimation of first-order and second-order moments and it is widely used in the literature for its fast convergence properties.} as the optimizer. Hyperparameter tuning is performed to determine the number of epochs, batch size, learning rate, and dropout rate. The train-test split is 75\% and 25\% with random partitioning.
Hence, the data set is first split into final train and test data sets with a 75:25 split. The final train set is further split in initial train and validation sets with again a 75:25 split.

Note that we segment the 
dataset into sessions so that the train-test randomly splits not samples but sessions. The output of the decoder is then averaged to obtain the typical MOS pattern. 

We also use the learned MOS pattern to categorize anomalous sessions. Note that the definition of \textit{anomaly} is a subjective matter. For example, having an average MOS less than 4, or having MOS lower than 3 in any part of the session can be considered an anomaly. All these anomaly definitions may have different repercussions and require different strategies in terms of network optimization, and thus, it is an area for further research. However, the ultimate goal of future networks is the autonomous identification of what constitutes an anomaly. In this respect, we consider an unsupervised learning approach wherein \textbf{the network learns typical application behavior and identifies those atypical instances as anomalies}.

\section{MOS Prediction}
\label{sec:MOS_Pred}
\noindent

In this section, we demonstrate that MOS can be predicted after appropriate enhancements of the available major network KPIs. 

\subsection{Preprocessing for MOS Prediction}
\noindent Note that our dataset has missing values in every row. Namely, the collections of measurement data are asynchronous. To remedy this, we fill the values in a backward manner for the inputs (i.e., the network KPIs). This ensures that our predictor does not invalidate causality in predictions, and each MOS value has a set of corresponding network KPI measurements. Note that one may also be inclined to fill for the intended output (i.e., MOS). However, the best method of such filling is not apparent, and a wrong choice may cause misleading results.

After handling the missing values, we also notice that the frequency of MOS measurements is much less than the network KPIs. Hence, it is also logical to include aggregated KPIs, in addition to the instantaneous KPI measurements corresponding to each available MOS value. We perform this aggregation in two different ways. Firstly, we aggregate by temporally averaging KPIs between successive MOS values so that the resulting aggregate can be added among the inputs to predict the superseding MOS value. Secondly, we employ a cumulative averaging that resets with each video session's start in our dataset. 
This operation is akin to some integrator in the discrete domain. For each session, and each KPI feature, an accumulator, which initializes at zero, is kept. When a feature is measured at some time during a session, the measurement is added to the corresponding feature accumulator for the said session. A counter, which keeps track of how many times a feature is measured during a session, is also kept alongside each accumulator. The cumulative average of a KPI feature in a session is output by dividing the corresponding accumulator by its pairing counter.

This is done to account for the accumulating effects occurring with playback buffer in adaptive video streaming.

\begin{figure}
	\centering
	\includegraphics[width=\linewidth]{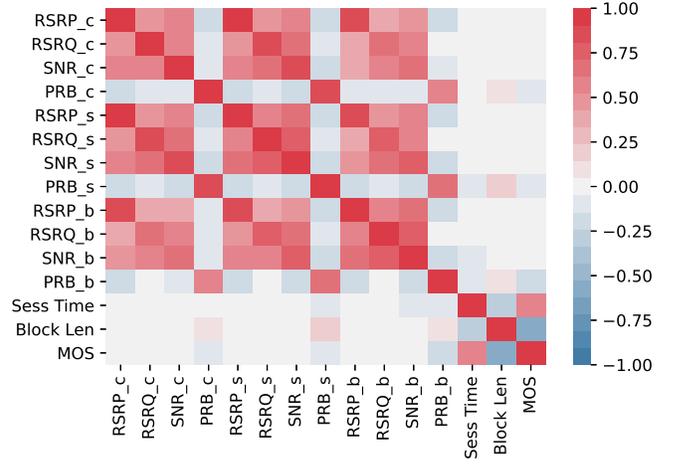}
	\caption{Input/Output Pearson Heatmap}
	\label{fig:pearson}
\end{figure}
\subsection{Additional Feature Enhancements}
\noindent As explained previously, MOS is also affected by application-specific buffering mechanisms. Furthermore, network KPIs may have additional delayed effects, and the temporally changing inherent qualities during a video session may also affect the user experience. Consequently, we add two temporal quantities among our set of inputs. Firstly, \textbf{the time difference, or the number of periodic network KPI measurements between successive MOS values}, is used as one of the predictors for, again, the superseding MOS value. In a testing scenario, this can be replaced with the size of a network measurement window, and the aggregates generated from KPIs between successive MOS values can be replaced with a moving average. Secondly, \textbf{the time since the start of a session or the total number of periodic network KPI measurements} is used, which, in a way, accounts for the MOS structure of the session. According to our findings, this predictor is the most dominant factor in MOS.


In Fig. \ref{fig:pearson}, we see how the enhanced KPI features and the MOS values are Pearson correlated, which is a form of variance-normalized co-variances. The features in this figure are explained as follows: 
\begin{itemize}
	\item `Sess Time' (session time) measures how much time has passed between the corresponding MOS value data-point and the start of its respective session.
	\item `Block Len' (block length) measures how much time has passed since the sampling of the previous MOS value until the corresponding MOS value.
	\item The subscripts $b$ and $s$ refer to the aggregates, such that $b$ stands for between successive MOS measurements, i.e., during the time `Block Len', and $s$ is used for the cumulative aggregation since the starts of respective sessions, for the duration of `Sess Time'. The subscript $c$ refers to the current or instantaneous measurements which are taken every second.
	\item `RSRP', `RSRQ', `SNR', and `PRB' are the network KPIs, or the RF measurement metrics utilized, as explained previously.
\end{itemize}

\subsection{Data Split and Learners}
\noindent After finalizing our input and output sets, we segment the samples into sessions so that the train-test split before actual learning will be the most distinct possibility. Specifically, not samples but sessions are randomly split into train and test sets. Then, the training is done using regression algorithms. In regression, the model outputs real values such as floating-point numbers based on input variables. We found ensemble models to be the best performing, and Random Forest Regression (RFR) and Gradient Boosting Regression (GBR) algorithms stood out, both of which utilized regression trees as base learners. Random Forest fits many decision trees to various subsets of the training data. Then, during prediction, it averages the individual predictions by these decision trees. In a way, RFR trains each decision tree on different data samples and, finally, instead of relying on a single tree, it merges all of them before taking the final decision. Meanwhile, Gradient Boosting also utilizes regression trees and builds an additive prediction model in a forward-stage fashion, where each subsequent predictor fits the remaining residuals. 

\begin{table}[]
\caption{MSE per session with different learners. \label{table:MSE_interpret}}
\centering
\begin{tabular}{lr}
\toprule
 Method & MSE per session \\
\midrule
Random Forest &           0.3850 \\
Gradient Boosting &           1.0176 \\
Decision Tree (all) &           1.0351 \\
Decision Tree (`Sess Time') &           1.2011 \\
\bottomrule
\end{tabular}
\end{table}
\subsection{Performance}
\noindent 
To evaluate the performance we use the coefficient of determination, i.e., $R^2$ score metric.
$R^2$ is computed as follows, where $y$ is the desired output and $\widehat{y}$ is its estimate.
\begin{equation*}
	R^2 = 1 - \frac{\mathbb{E}[(y - \widehat{y})]}{\text{Var}(y)}.
\end{equation*}
Given that we have $N$ number of data points, i.e., $y_i$, $i=1,\ldots, N$ and $N$ corresponding estimates $\widehat{y}_i$, $i=1,\ldots, N$, this metric can also be interpreted as follows:
\begin{equation*}
	R^2 = 1 - \frac{\sum_{i=1}^N (y_i - \widehat{y}_i)^2}{\sum_{i=1}^{N} (y_i - \overline{y})^2}.
\end{equation*}
In effect, we compare our estimator with the one which outputs fixed-point $\widehat{y}$, via the residual sum of squares. Perfect estimator with $\widehat{y}_i = y_i$ results in the maximum possible value $R^2=1$, while a baseline fixed-point estimator of $\widehat{y}_i = \widehat{y}$ for all $i$ results in $R^2=0$. $R^2$ can also be negative for estimators that work worse than the fixed-point baseline estimator in the least-squares sense.

We performed regression with RFR and GBR with 75\% train-test data split over 50 random trials.
As said before, the split is done over sessions. However, the estimations are done in a per-sample basis for each test session.
The mean of the coefficient of determination ($R^2$) over test data for RFR and GBR are $0.60676$ and $0.62837$, respectively. The 95\% confidence intervals of ($R^2$) for RFR and GBR are $(0.59859, 0.61493)$ and $(0.61963, 0.63711)$, respectively.



We have calculated the mean square error (MSE) per session of predicted MOS averaged over all sessions for different learners as given in Table \ref{table:MSE_interpret}. We also considered decision tree regression with two different sets of input features. Firstly, we consider a decision tree regression trained over the same set of features as RFR and GBR. Secondly, \textit{only} `Sess Time' is used as the single input feature for another decision tree regression. Interestingly, the second tree's MSE per session is not far from that of the first decision tree. This is because `Sess Time' is the most dominant predictor. However, we will show in the next section that `Sess Time' is not sufficient to accurately predict MOS during anomalies.

\begin{figure*}
    \centering
    \includegraphics[width=1\textwidth]{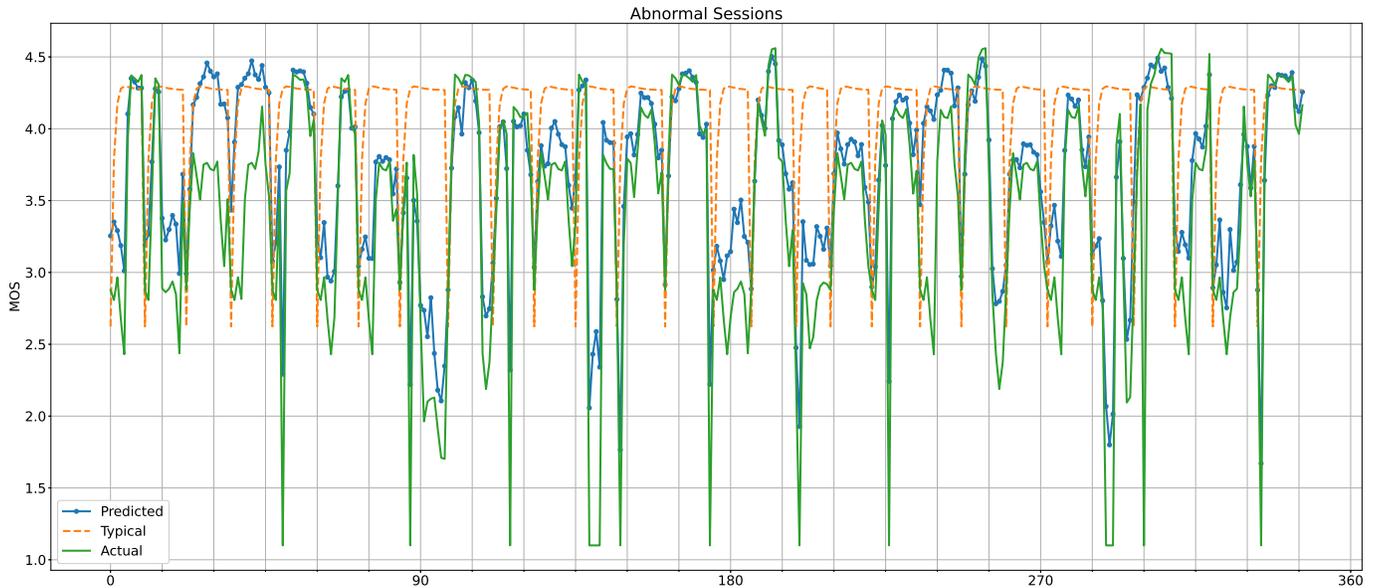}
    \caption{Expected, actual, and predicted MOS patterns for 
	multiple
 anomalous video test sessions.}
    \label{fig:abnormal}
\end{figure*}
\begin{figure}
    \centering
    \includegraphics[width=0.4\textwidth]{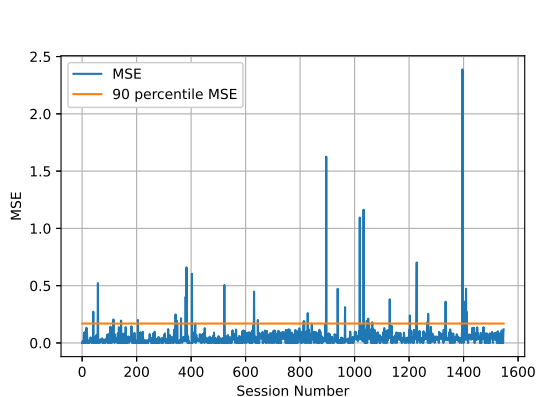}
    \caption{MSE of MOS prediction of benchmark video sessions as compared to the typical MOS pattern.}
    \label{fig:MSE}
\end{figure}
\begin{figure}
    \centering
    \includegraphics[width=0.4\textwidth]{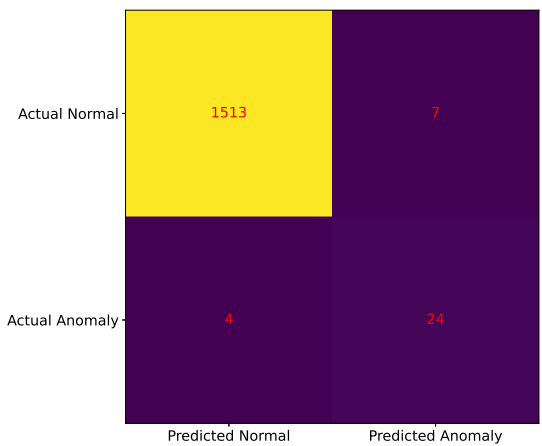}
    \caption{A sample confusion matrix.}
    \label{fig:confusion}
\end{figure}
\begin{figure}
    \centering
    \includegraphics[width=0.4\textwidth]{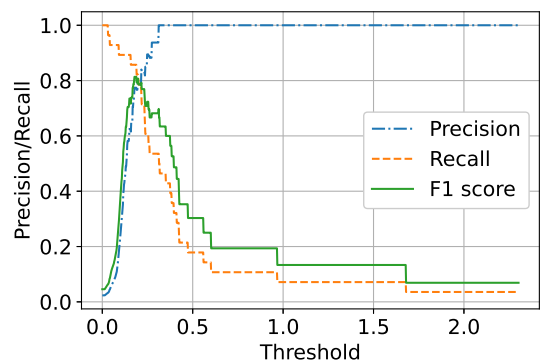}
    \caption{Precision-Recall and F1 values.}
    \label{fig:precision_recall}
\end{figure}
\section{Anomaly Detection}
\label{sec:anomaly}
\noindent Recall that the main challenge in network performance management is the unavailability of user feedback during regular operation. Consequently, actual user satisfaction is unknown on the network side. For this reason, we use the RFR predictor explained in the previous section that can predict user MOS based only on network KPIs. In Fig. \ref{fig:abnormal}, we demonstrate predicted, typical, and actual MOS patterns over the video sessions identified as anomalous by using the RNN described in Fig.\ref{fig:NN_typical_MOS}. 
Fig. \ref{fig:abnormal} displays a total of 28 sessions in succession.

Note that although the predicted MOS values do not always match the actual MOS values, the predicted values follow the actual MOS measurements quite closely. Hence, we use the predicted MOS values as a proxy of actual MOS measurements. 

Next, we calculate the MSE of each session by calculating the error as the difference between \textbf{the predicted MOS} and the \textbf{\textit{typical} MOS} values. Fig. \ref{fig:MSE} shows the MSE values of all sessions, and the solid red horizontal line shows the value of MSE for the 90 percentile. If a session has an MSE higher than a given threshold, the session is deemed anomalous. 


Fig. \ref{fig:confusion} shows the confusion matrix for a sample output for our anomaly classifier. The confusion matrix allows for easy visualization of mislabeling of actual and predicted normal and abnormal sessions. In this figure, \textit{actual} abnormal sessions are those sessions identified in the same way as explained before but this time error is calculated as the difference between \textbf{the true MOS} and typical MOS values for each session. Accordingly, 28 out of 1199 video sessions in our dataset are categorized as 
actually anomalous. In this sample output, using RFR based MOS prediction, 7 anomalies are missed and 4 normal sessions are mislabeled as anomalies.

In order to better understand the performance of our anomaly detection model, we calculate the precision, recall, and F1 score metrics. Note that accuracy is not chosen as a performance metric. This metric is only applicable when the dataset is symmetric, which is clearly not the case in our dataset, with anomalies being rare events. 
In classification problems, true positives and negatives are the observations that are correctly predicted. Meanwhile, false positives and negatives occur when the actual class contradicts the predicted class. \textit{Precision} is defined as the ratio of correctly predicted positive observations to the total predicted positive observations. Meanwhile, \textit{recall} is the ratio of correctly predicted positive observations to all of the actual positive observations. Finally, \textit{the F1 score} is the harmonic mean of precision and recall. Fig. \ref{fig:precision_recall} shows a sample output of our anomaly detection model. Threshold refers to the minimum MSE per session over which the session is classified as an anomaly. We performed 50 experiments with random train-test splits for RFR MOS prediction. For each experiment, we determine the maximum F1 score for the anomaly detector. The maximum F1 score is averaged over these experiments, and it is calculated as $0.77470$ with a 95\% confidence interval of $(0.76217, 0.787234)$. 


\begin{figure}
    \centering
    \includegraphics[width=0.5\textwidth]{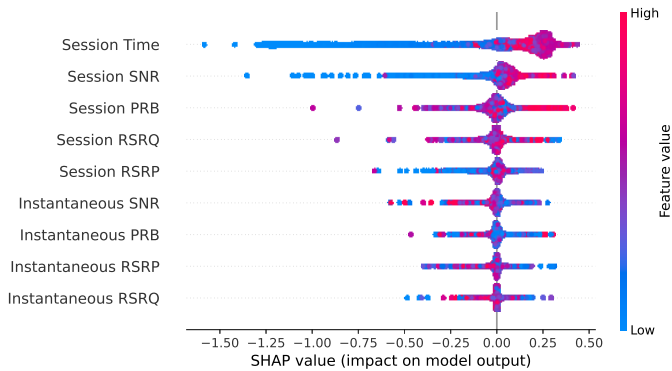}
    \caption{SHapley Additive exPlanations of the features used in the model.}
    \label{fig:shapley_visual}
\end{figure}

\begin{figure*}
\begin{subfigure}{1.0\textwidth}
 \centering
        \includegraphics[width=0.6\textwidth]{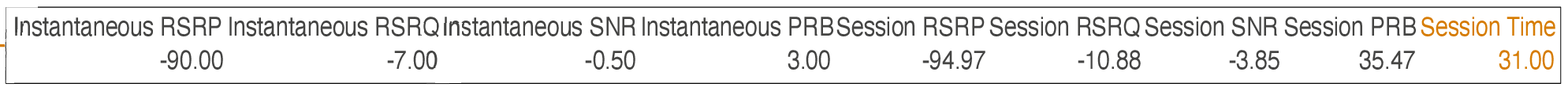}
\end{subfigure}

\begin{subfigure}{1.0\textwidth}
 \centering
        \includegraphics[width=0.65\textwidth]{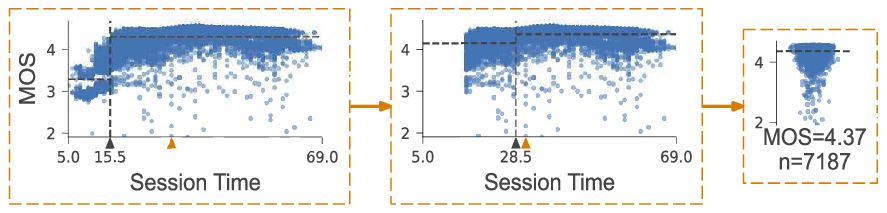}
\caption{MOS prediction by a single feature: Session time.}
\end{subfigure}

\vspace{5mm}

\begin{subfigure}{1.0\textwidth}
   \centering
        \includegraphics[width=0.8\textwidth]{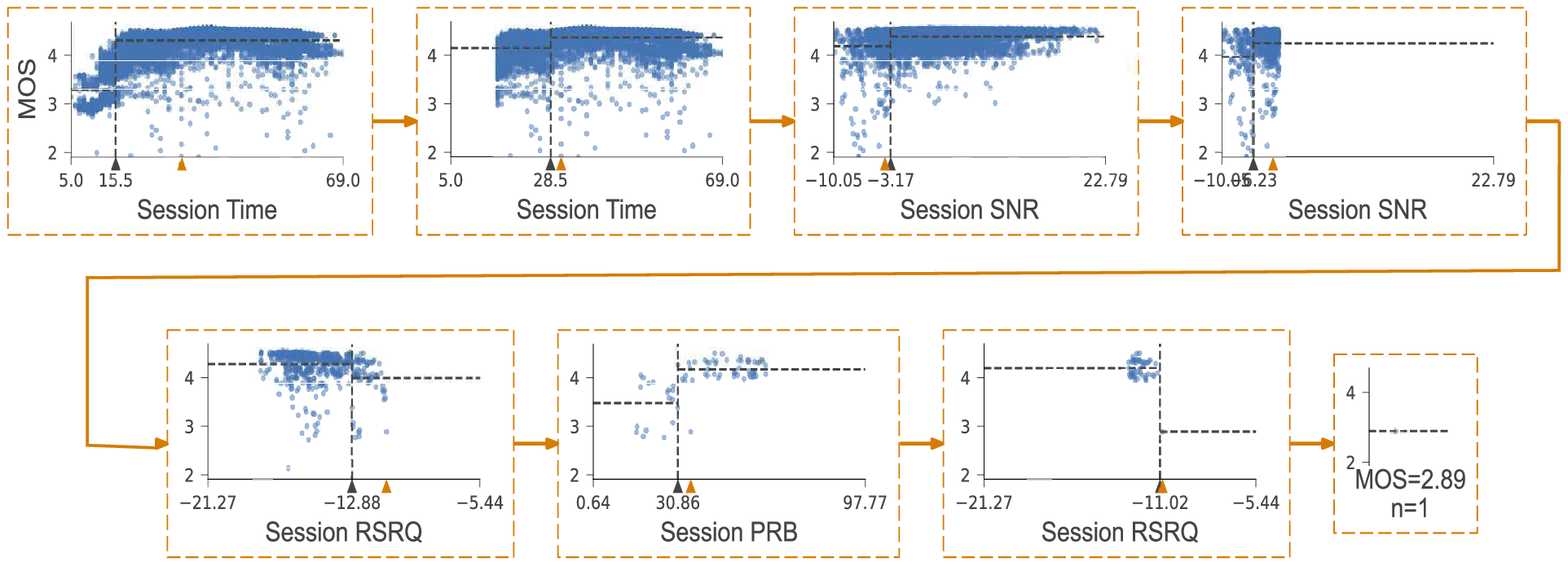}
    \caption{MOS prediction with all features combined. The maximum depth is limited to 8.}
\end{subfigure}
\caption{The actual MOS value is $2.43$ and the session time is 31 seconds into a session of 60 seconds. }
\label{fig:MOS_time_effect_less}
\end{figure*}

\begin{figure*}
\begin{subfigure}{1.0\textwidth}
 \centering
        \includegraphics[width=0.6\textwidth]{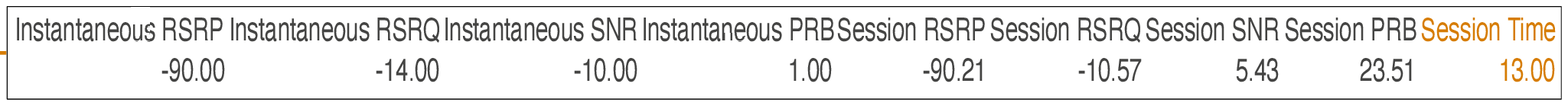}
\end{subfigure}

\begin{subfigure}{1.0\textwidth}
 \centering
        \includegraphics[width=0.7\textwidth]{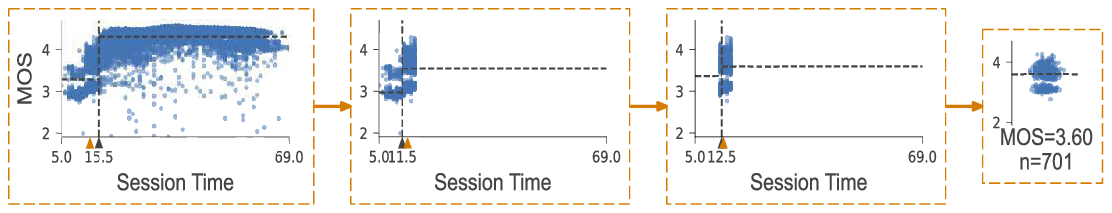}
\caption{MOS prediction by a single feature: Session time.}
\end{subfigure}

\vspace{5mm}

\begin{subfigure}{1.0\textwidth}
   \centering
        \includegraphics[width=0.8\textwidth]{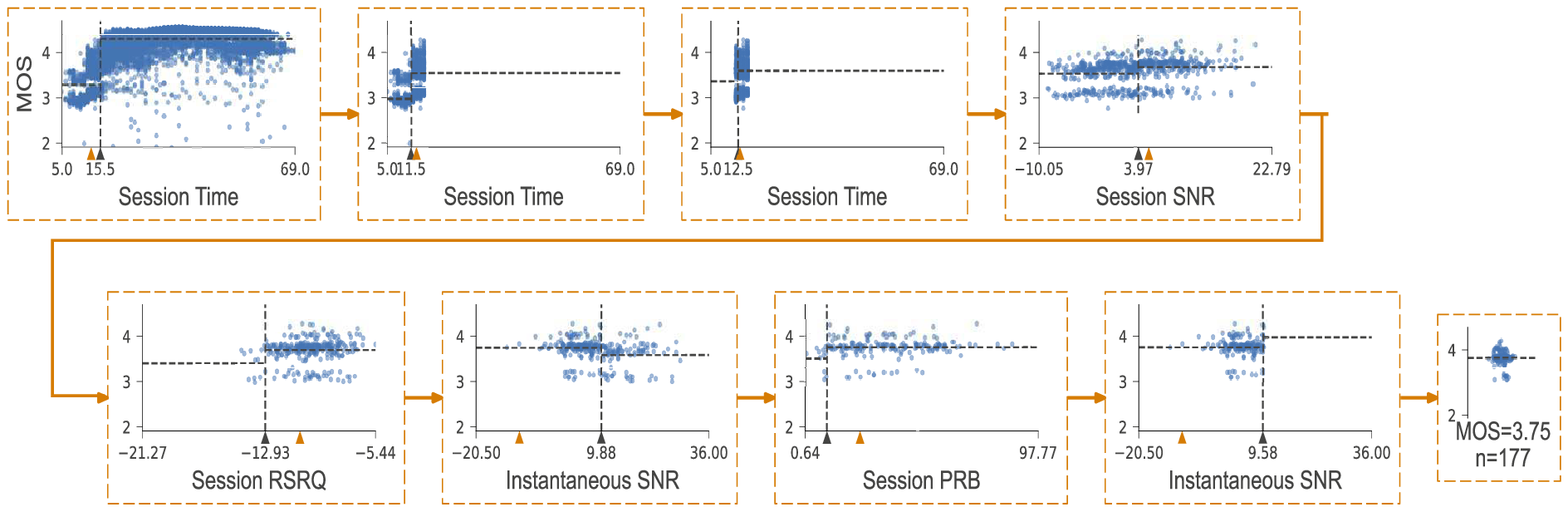}
    \caption{MOS prediction with all features combined. The maximum depth is limited to 8.}
\end{subfigure}
\caption{The actual MOS measurement is $3.819$ and the session time is 13 seconds into a session of 60 seconds. }
\label{fig:MOS_time_effect_more}
\end{figure*}
\section{Interpretation of ML models}
\label{sec:interpret}
\noindent In this section, we aim to interpret the output of our proposed ML models. Recall that our proposed MOS prediction method uses ensemble learning to achieve an accurate prediction of MOS values based on the radio network KPIs. Ensemble learning is usually treated as a black box, which in turn is not interpretable. On the other hand, decision tree learning is considered a white box since it is intuitively clear that there is a path from the root of the tree to the leaf for each decision that a tree makes. The decision path consists of a series of decisions, each guided by a particular feature and contributing to the final prediction. Therefore, we use decision trees to understand the decisions of our RFR model. In order to do this, we first train the decision tree regressor with the output of our RFR model for different sets of input features. Although decision tree regression trained with RFR outputs will give worse performance than the original RFR, it will allow us to follow the decision paths to MOS prediction.


In this section, we use more easily identifiable names for the features of the ML model described in Section \ref{sec:MOS_Pred}. Namely, \textit{Session} (SNR, PRB, RSRQ, RSRP) refer to the features (`SNR', `PRB', `RSRQ', `RSRP') with subscript $s$; and \textit{Instantaneous} (SNR, PRB, RSRQ, RSRP) refer to the features (`SNR', `PRB', `RSRQ', `RSRP') with subscript $c$. Finally, \textit{Session Time} refers to `Sess Time'. 

We first analyze the feature importance by plotting SHAP (SHapley Additive exPlanations) values for our decision tree model in Fig.\ref{fig:shapley_visual}. SHAP values are used in the literature to understand a feature's contribution to a change in the output of the model. As the name implies, SHAP takes into account Shapley values from game theory, where the Shapley value is defined as the average expected marginal contribution of one feature after all possible combinations have been considered \cite{NIPS2017_7062}. As Fig.\ref{fig:shapley_visual} reveals, Session Time is the most important feature in predicting the MOS, and Session SNR comes in a distant second. The rest of the features are listed in the order of their importance level, but their combined effect on the predicted MOS is relatively small. However, as exhibited earlier, even though the features with low SHAP values do not improve the model output on average, they are nevertheless critical in detecting anomalies. A particular example is the Session PRB. When this feature has a positive contribution to the model output only when it has a high (red) value.



Although session time stands out as the most important feature, a decision tree that uses session time as its single input feature fails to predict well during the anomalies, as demonstrated in the following examples. In Fig. \ref{fig:MOS_time_effect_less}, and in Fig. \ref{fig:MOS_time_effect_more}, we show the path of the decision tree for a particular MOS prediction in the corresponding video sessions. The decision path indicates the sequence of nodes in the decision tree that is followed until the final prediction. Each box depicted in the figures corresponds to a node in the decision tree and shows the distribution of MOS measurements with respect to the feature used for decision at that node. The same figure also displays the threshold used to split the tree with a small red arrow indicating the threshold for further splitting. 

In particular, in Fig. \ref{fig:MOS_time_effect_less} (a) and (b), we consider a video session with an actual MOS value of 2.43 at 31 seconds into the session. In Fig. \ref{fig:MOS_time_effect_less} (b), the first feature used to split the data is the session time with a threshold of 15.5s. Hence, if we are predicting the MOS at 31s, we fall into the right-hand side of this threshold. The second node also uses the session time but this time with a threshold of 28.5s. The third node of the decision tree is session SNR, which is the average of SNR over the first 31s, with a threshold \mbox{-3.17dB}. Since our session SNR is \mbox{-3.85dB}, we can further investigate all measurements to the left of this threshold. The subsequent features used to narrow down the prediction are session SNR (again), session RSRQ, session PRB, and session RSRQ again resulting in a final prediction of a MOS value of 2.89. 

Meanwhile, Fig. \ref{fig:MOS_time_effect_less} (a) shows the decision path of a tree which only uses session time as its input feature. As discussed earlier, in a typical video session, after an initial short time, the MOS value settles around 4.25, which is what this tree approximately predicts. However, this prediction is far away from the actual measurement. This is due to the session possibly experiencing atypical RF conditions. 

In the second example, we consider a session with an actual MOS equal to 3.819 at 13 seconds into the said session. At this time instant, the video has just come off a steep MOS increase by reaching its steady-state throughput, reaching a MOS of 4.2 (approximately). Fig. \ref{fig:MOS_time_effect_more} (a) shows the decision path of a tree using only session time, giving a respectable MOS prediction of 3.60. Using all features only improves the MOS to 3.77 since the session possibly experienced typical RF conditions.

In Fig. \ref{fig:MOS_time_effect_less} and \ref{fig:MOS_time_effect_more}, we demonstrate the relative effect of time into the session over two examples. In Fig. \ref{fig:MOS_time_effect_less}, the session time is 31 seconds, and if MOS prediction is made with only this feature MOS prediction would be 4.37 whereas the actual measurement is 2.43. The discrepancy is because in most of the sessions, at this time instant, the MOS is usually high due to the buffering of video frames. However, in this particular session, the radio network KPIs were not satisfactory and the buffering was not sufficient to reach this MOS level. Such a discrepancy is identified by taking into account radio KPIs correctly predicting MOS. Meanwhile, in Fig. \ref{fig:MOS_time_effect_more}, MOS is predicted for a session time of 13 seconds into the session. Since we are still early into the session, buffering is not yet the dominant factor in the quality of the video. Hence, session time has a more prominent effect on the prediction of MOS, i.e., by only session time MOS is predicted as 3.60 when the actual measurement is 3.819. Adding radio network KPIs improves the prediction to 3.77.

\begin{figure}
    \centering
    \includegraphics[width=0.5\textwidth]{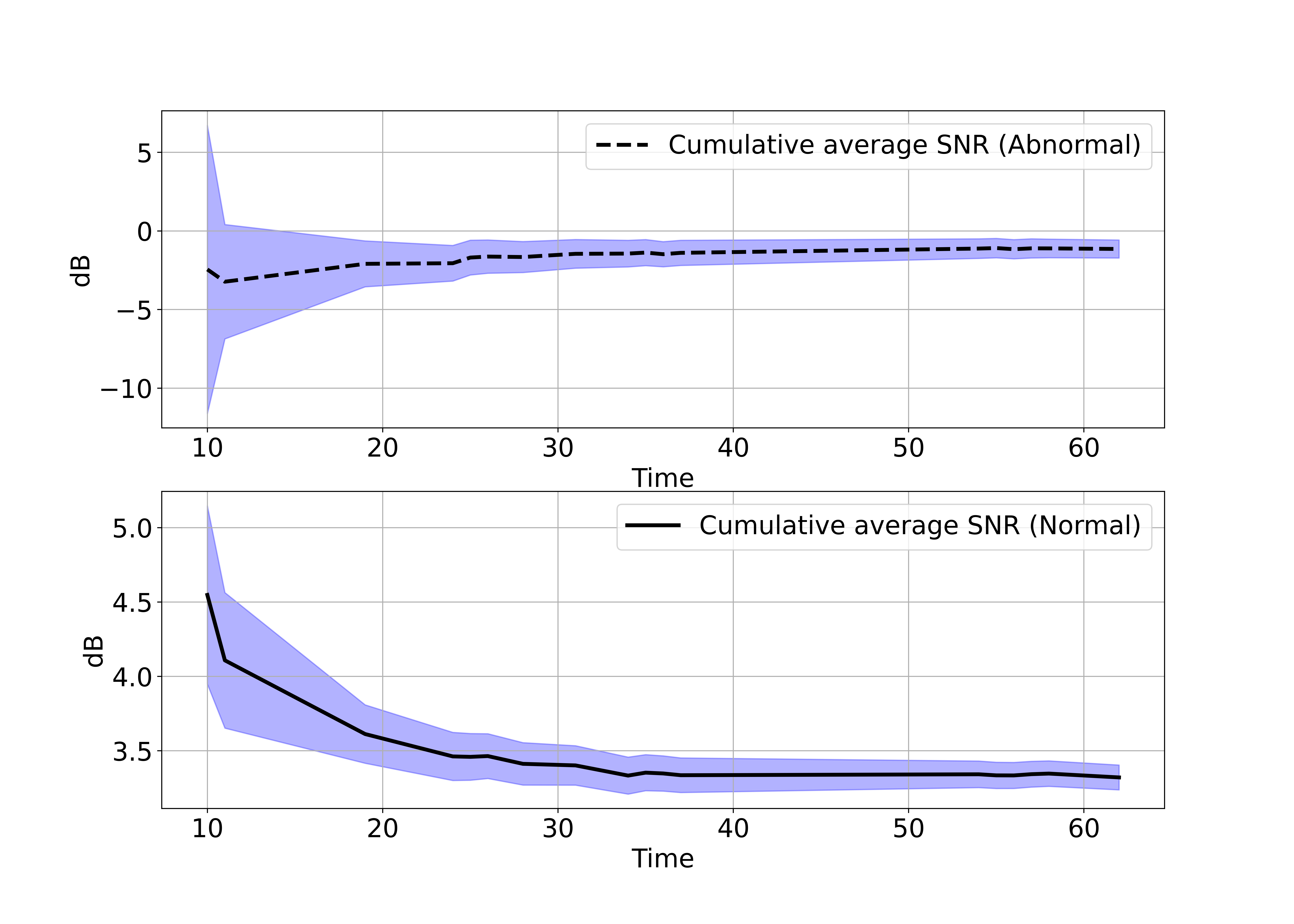}
    \caption{Cumulative Average of SNR for normal and abnormal sessions with respect to time. The shaded region shows the 95\% confidence interval.}
    \label{fig:root}
\end{figure}
\section{Root Cause Analysis}
\label{sec:root}
\noindent It is easy to envisage the anomalies are due to the deep fades in the wireless channel. In this section, we check the correctness of this intuition by verifying it through the output of our proposed anomaly detection method. Fig. \ref{fig:root} depicts the cumulative average of SNR for normal and abnormal sessions. The cumulative average SNR is calculated by averaging the value of feature `$SNR_s$' from the beginning of the session until the desired time into the session. The shaded area represents the 95\% confidence interval. Note that the cumulative average SNR for abnormal sessions varies between -4dB and -0.5dB, and it changes between 3.3dB and 4.5dB for normal sessions. Hence, as expected, cumulative average SNR is a good indicator to detect anomalies in the video session. However, its variation is initially very high, and in fact, at the first 10 seconds into the session, both normal and abnormal sessions have similar SNR statistics. The cumulative average SNR becomes an accurate indicator further into the session due to the averaging, which reduces the said variation. 


\section{Conclusions and Future Directions}
\label{sec:conc}
In this paper, we developed an autonomous QoE prediction and anomaly detection architecture that complements real drive-tests. 
By using our proposed architecture, the operators will be able to evaluate QoE at different geographical locations and at any time by using base station traces. Principally, the proposed architecture has three main components:
\begin{enumerate*}[label=\roman*)]
    \item a pattern recognition component that learns typical application behavior;
    \item a predictor that learns QoE value from network KPIs;
    \item a component that compares the predicted QoE metric values with the typical application behavior to detect anomalies.
\end{enumerate*}
Although our architecture is sufficiently general, we demonstrate its performance with the YouTube application, since QoE metric is well defined for adaptive video streaming, and YouTube constitutes a large data volume for cellular networks.

Network automation is becoming one of the key enablers for 5G and beyond networks, and many standard development organizations (SDOs) such as International Telecommunication Union (ITU) focus group on Autonomous Networks, 3GPP Network Data Analytics Function (NWDAF), and ETSI Experiential Networked Intelligence (ENI), are addressing standardization gaps in this area ranging from data collection and data format definition to architecture. These SDOs would benefit from contributions involving requirements for implementing service-based QoE estimation using the measurements of the network QoS metrics, which also consider distributed architectures for reduced estimation delays and minimal additional backhaul traffic.

When regarding service-based QoE prediction and its latter operations in a network operator perspective, there might be an opportunity to extend this work by developing and adapting it for different network services. In this context, Fixed Wireless Access (FWA), rising as an emerging and promising technology in recent years, particularly with the advent of LTE Advanced and then 5G, seems to be an appropriate candidate for future work. Further, a location-based QoE model might be valuable since FWA products (i.e., Customer-premises Equipment, CPEs) are used in fixed locations according to the subscription agreement of this service. 

In addition, the automated network optimization solutions (e.g., Self-Organizing Network, SON), currently utilized by the major network operators in the world, basically work based on cell-based network KPIs that reflect any degradation or change in the voice and data services. However, it might be a chance to manage these SON platforms more competently by predicting user experience at the application level via a QoE model enriched with AI/ML techniques, instead of having a traditional triggering mechanism depending on the aggregated network KPIs.

\bibliographystyle{bib/IEEEtran}
\balance\bibliography{bib/IEEEabrv,bib/IEEEexample,bib/bib_2}

\end{document}